\documentclass[%
 reprint,
 amsmath,amssymb,
 aps,
 prl,
]{revtex4-1}
\usepackage{xcolor}
\makeatletter
\usepackage{graphicx}
\usepackage{dcolumn}
\usepackage{bm}
\usepackage{float}
\usepackage{gensymb,comment}
\usepackage{soul}
\usepackage{ulem,lineno}
  \usepackage{hyperref}
    \usepackage{url}

\renewcommand{\sout}[1]{}

\colorlet{red}{black}

\begin{document}

\title{Typicality of the 2021 Western North America Summer Heatwave} 

\author{Valerio Lucarini}%
\affiliation{Department of Mathematics and Statistics \& Centre for the Mathematics of Planet Earth, University of Reading, UK
}
\email{Corresponding author. Email: v.lucarini@reading.ac.uk}

\author{Vera Melinda Galfi}
\affiliation{Department of Earth Sciences, Uppsala University, Uppsala, Sweden}
%

\author{Gabriele Messori}
\affiliation{Department of Earth Sciences and Centre of Natural Hazards and Disaster Science (CNDS), Uppsala University, Uppsala, Sweden \& Department of Meteorology and Bolin Centre for Climate Research, Stockholm University, Stockholm, Sweden}

\author{Jacopo Riboldi}
\affiliation{Department of Earth Sciences, Uppsala University, Uppsala, Sweden}

\date{\today}

\begin{abstract}
Elucidating the statistical properties of extreme meteo-climatic events and capturing the physical processes responsible for their occurrence are key steps for improving our understanding of climate variability and climate change and for better evaluating the associated hazards. {\color{red}It has recently become apparent that large deviation theory is very useful for investigating persistent extreme events, and specifically, for flexibly estimating long return periods and for introducing a notion of dynamical typicality. Using a methodological framework based on large deviation theory and taking advantage of long simulations by a state-of-the-art Earth System Model, we investigate  the 2021 North America Heatwave.}  Indeed, our analysis shows that the 2021 event \sout{is not an exception - a \textit{Dragon King} - and }can be seen as an unlikely but possible manifestation of  climate variability, whilst its probability of occurrence is greatly amplified by the ongoing climate change.  We also clarify the properties of spatial coherence of the 2021 heatwave and elucidate the role played by the Rocky Mountains in modulating hot, dry, and persistent extreme events in the Western Pacific region of North America.

\end{abstract}

\maketitle


\section{Introduction}
Investigating the statistical properties of extreme meteo-climatic events and the physical mechanisms behind their occurrence is an area of ever-growing interest in climate science \cite{Ipcc2012}. \sout{On the one hand, meteo-climatic extremes are the natural hazards leading to arguably the most relevant risks to human and environmental welfare. On the other hand}{\color{red}Additionally, their study allows to access some key aspects of the climate system, since there is in general a fundamental connection between extreme events and the dynamics of the system generating them }\sout{the study of extreme events allows one to investigate some key aspects of the climate system, as there is a fundamental connection between extreme events and dynamical features of the system under investigation }\cite{Lucarini2016extremes,Galfi2017,Faranda2017,Hochman2019,Galfi2021}. Long-lasting meteo-climatic extremes are hazards of particular relevance because anomalous and persistent conditions are conducive to extremely damaging impacts on {\color{red}ecosystems and society} \sout{natural and engineered systems} \cite{Easterling2000,Who2004,Poumadere2005,Ipcc2012,wang2018quantitative}. Moreover, atmospheric phenomena with a long characteristic time scale are usually also characterised by a large spatial extent \cite{Ghil2002,Ghil2020}, potentially leading to systemic risk \cite{Kornhuber2020}. {\color{red}In general, the occurrence of extremes of surprisingly large magnitude might be indicative  of an approaching critical transition \cite{Faranda2014,Franzke2022}}. \sout{In this sense, unusual climatic extremes (sometimes referred to as \textit{$5$-$\sigma$ events}) might be responsible for triggering tipping behaviour in the Earth system.}



\begin{figure}
\includegraphics[width=0.8\linewidth]{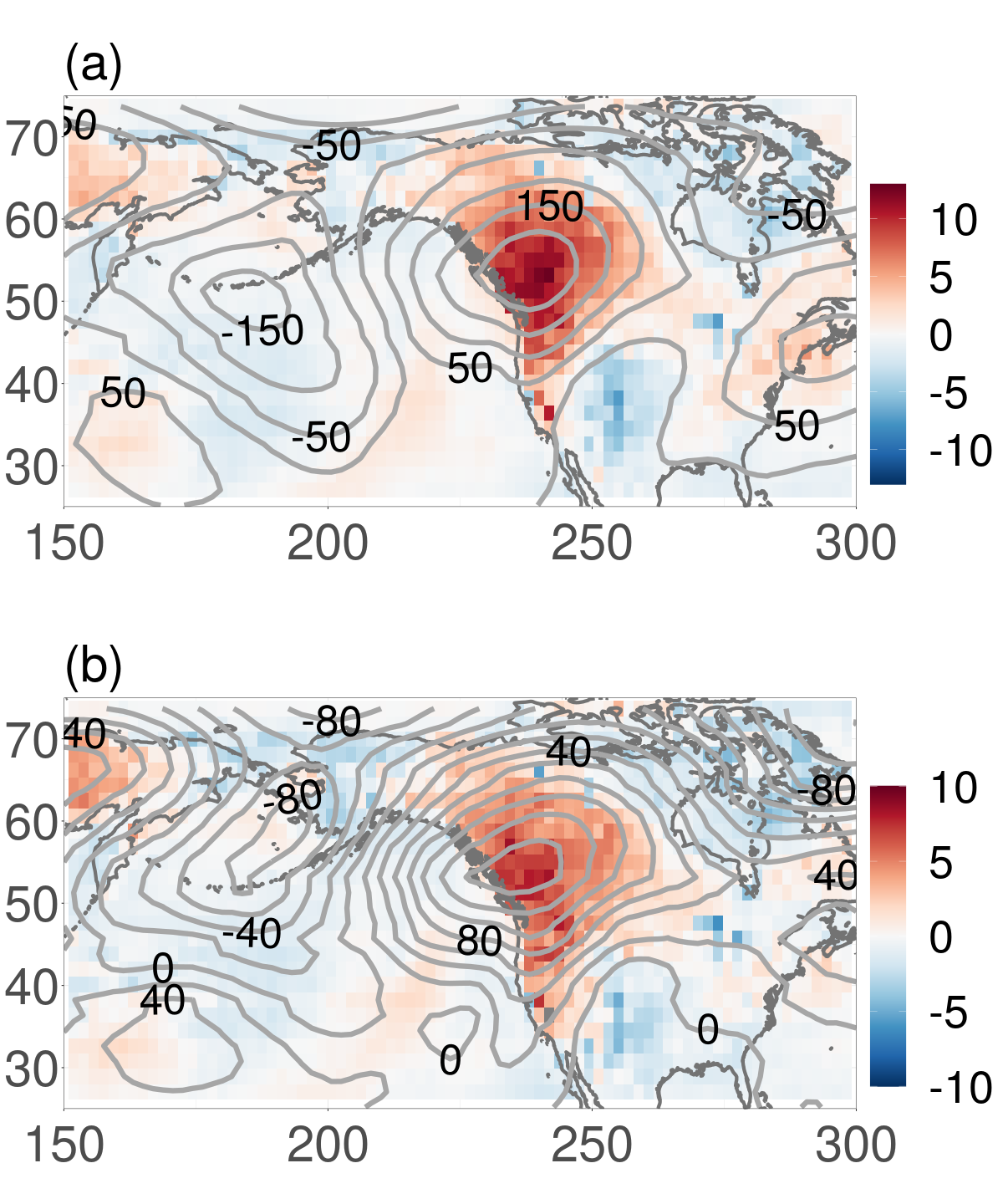}
\caption{ERA5 anomalies for (a) June 24th - June 30th 2021 and (b) June 24th - July 8th 2021 average of {\color{red}T2M} (colours, in K) and Z500 (contour lines, in m) with respect to the 1991-2020 climatology for the corresponding period.\label{anomaly7}}
\end{figure}

\paragraph{The 2021 Western North America Heatwave}
Between late June and mid-July 2021 a large latitudinal band of the the western sector of North America faced a very intense heatwave, in the context of the ongoing megadrought in the region \cite{Zhang2021,Zhang2022}. Several new temperature records were established: $49.6^\circ$C were measured in Lytton, which is the new absolute temperature record for  2-m temperature {\color{red}(T2M)} for Canada. It is estimated that the heatwave has led to over 1000 directly attributable deaths and damages worth several billion USD, associated with wildfires, crop loss, infrastructural damages, and flooding due to rapid snow melt \cite{Philip2021,Thompson2022,Lin2022}. \sout{; see \cite{Philip2021,Thompson2022,Lin2022} and references therein.}  The record-breaking temperatures were the result of preconditioning combined with an anomalous atmospheric circulation. Prior to the heatwave, the southern portion of the region was anomalously dry \cite{Philip2021,Osman2022} and it is known that long-term drought increases the probability of occurrence of heatwaves, as a result of the drying of the soil \cite{Zampieri2009,Miralles2014,Schumacher2019}, possibly leading to the onset of a cycle of heat, drought, and fire \cite{Sutanto2020,Libonati2022}. This was compounded by the presence of an intense atmospheric $\Omega$-block,  like the one associated with the 2010 Russian heatwave \cite{Dole2011,Lau2012,Galfi2021PRL}.  Atmospheric blocks \cite{Woolings2018} are often the culprits for individual heatwaves as a result of possibly concurring processes, such as anomalies in clear-sky incoming radiation, warm air advection, and subsidence   \cite{Jimenez2022,Chan2022}. 
The {\color{red}Western North America Heatwave (WNAHW)} blocking developed on June 25th in association with a large scale pattern over the mid-latitudes dominated by low zonal wavenumbers \cite{Thompson2022,Bartusek2022}. The ridge reached an impressive maximum 500 hPa geopotential height (Z500) of almost 6000 m over western North America. Figure \ref{anomaly7} shows the ERA5 \cite{ERA5} {\color{red}T2M} and Z500 anomaly fields averaged over periods of 7 and 15 days {\color{red}encompassing} \sout{centred around} the heatwave.  
The agreement between {\color{red}the large scale patterns} in Figs. \ref{anomaly7}a)-b) clearly shows the persistent and {\color{red}spatially coherent} nature of the event. 
Predicting the onset of blocking is  extremely challenging  for numerical models \cite{Pelly2003,matsueda2018estimates,LucariniGritsun2020} and, indeed, very few if any of the weather forecasting systems were able to estimate accurately the intensity of the 2021 WNAHW  at lead times of a few weeks \cite{Lin2022}.

\sout{Very r}Recently, a team from the World Weather Attribution initiative has concluded, using a multi-model and multi-method attribution analysis applied to daily maximum temperatures, that it is possible to attribute almost non-ambiguously the 2021 event to climate change, because its likelihood in present climate conditions (factual world) is estimated to be orders of magnitude higher than in past climate conditions (counterfactual world) \cite{Philip2021}, according to the methodology presented in \cite{Hannart2016}. 

\paragraph{This Study}

In \cite{Galfi2021PRL,Galfi2021} it was shown that large deviation theory (LDT) \cite{Varadhan1984,Touchette2009} allows for constructing a  climatology of heatwaves and cold spells over land. This approach is  useful because one can go beyond purely descriptive statistics. Indeed, the return times of sufficiently persistent and intense temperature anomalies can be predicted by constructing the so-called rate functions for each given spatial location. For temperature timeseries, such rate functions are to a good approximation quadratic. One can thus construct them, in close agreement with the central limit theorem, by using relatively simple statistics like the  variability and integrated autocorrelation time of the daily temperature records.

\sout{Our intuition might find it hard to associate the occurrence of rare events to the notion of typicality.} 
It has recently become apparent that LDT can also be used to introduce a {\color{red}- somewhat counter-intuitive -} notion of  typicality of large and persistent temperature fluctuations associated with extreme heatwaves \cite{Ragone2017,RagoneBouchet2021,Galfi2021PRL} or cold spells \cite{Galfi2021PRL,Galfi2021}. 
{\color{red}The intense and persistent temperature fluctuations observed in a specific location are associated with a very unusual, rarely seen large-scale atmospheric configuration. Yet, if we subsample our dataset conditional on the occurrence of comparable temperature fluctuations in the same location, such large scale patterns are similar to each other, and are, in the sense discussed below, \textit{typical}.}
\sout{Given a  target region of interest, the large and persistent fluctuations we observe will take place, apart from small-scale spatio-temporal fluctuations, in association with a specific large-scale atmospheric configuration. Intuitively, this configuration should be very rare according to the natural variability. However, conditional on the occurrence of an extreme temporal fluctuation at the target location, such a configuration may be typical. In other words, the probability of the occurrence of a large scale atmospheric pattern that is very similar to the one observed can be very high given the occurrence of an extreme heatwave or cold spell, and gets closer to one as we consider more stringent criteria - in terms of intensity and duration - for the temperature fluctuation. Such a concept of typicality does not exclude the possibility of exceptional events (e.g. \textit{Dragon Kings} \cite{Sornette2012}), but these are much less likely than the typical (yet very rare) extreme events. This study \sout{also} {\color{red}further} presents a more detailed and critical account of the methodology presented in \cite{Galfi2021PRL}, which supports its robustness.}  

The main goal of this study is to investigate whether the \sout{2021 Western  North America heatwave (}2021 WNAHW\sout{)} is dynamically typical according to the LDT viewpoint presented in \cite{Galfi2021PRL,Galfi2021}, \sout{or is, instead, a \textit{Dragon King} that should be interpreted as a climatic surprise \cite{GritsunLucarini2017}.}\textit{i.e.} whether the corresponding large-scale circulation pattern can be considered to belong to the subset of typical circulation patterns mentioned above. We will also try to better understand the geographical features of the 2021 WNAHW by elucidating its spatial coherence. This study also presents a more detailed and critical account of the methodology presented in \cite{Galfi2021PRL}, which supports its robustness. 

Note that, through the use of an Earth System Model run in preindustrial, steady state configuration as in \cite{Galfi2021PRL}, we will study the  event as a fluctuation with respect to a baseline climatology. Hence, our approach is \sout{for the most part }different from that of the attribution studies mentioned before, where what matters is the absolute temperature reading, rather than the anomalies. We will estimate the impact of shifting the climatology in order to roughly account for the impact of historical climate change on the probability of occurrence of events comparable to the 2021 WNAHW. {\color{red}While this is obviously a serious simplification, because one neglects all the possible impacts of shifting climate conditions on climate variability (e.g. changes in the dynamics of the atmosphere, water cycle, soil properties \cite{Chan2022}), shifting the climatic mean has been shown to be a good first order approximation of the impact of climate change on the statistics of heatwaves \cite{Thompson2022}. }

\sout{We remark that t}The 2021 WNAHW featured a particularly acute phase between June 27$^{th}$ and June 29$^{th}$. \sout{As a result, }Hence, some studies have focused on the statistical investigation of daily temperature records \cite{Philip2021,Thompson2022} and taken advantage of the statistical framework provided by extreme value theory \cite{Coles2001,Lucarini2016extremes}. \sout{Instead, s}Since persistence plays a major role in determining the impact of heatwaves \cite{Poumadere2005}, we prefer to look into cumulative temperature anomalies. 

A succinct \sout{yet self-contained} account of the relevant LDT background that supports our analysis is presented in the next section. This is then followed by the discussion of the main findings of this investigation. A section presenting the concluding remarks and the perspectives for future investigation concludes this contribution. This letter is accompanied by {\color{red}supplementary material (SM)\cite{Lucarini2022_SM}} that complements the core findings described here.

\section{The Mathematical Background}
We follow below the viewpoint developed for fluid flows in \cite{Dematteis2019,DematteisSIAM2019}, which is based upon the theoretical framework presented in \cite{Touchette2015}. We consider a continuous-time chaotic dynamical system defined by an ordinary differential equation of the form
\begin{equation}
    \dot{x}=G\left(x\right)
\end{equation}
with $x\in\mathbb{R}^N$ with initial condition $x(0)=X$ belonging to the attractor $\Omega$ of the system. {\color{red}The solution at time $t\geq0$ is given by $x(t,X)=S^tX$ where $S^t$ is the evolution operator up to time $t$.} We assume that the system possesses a unique physical invariant measure $\rho$ supported on $\Omega$ 
and that our initial condition $X\in\Omega$, \textit{i.e.} 
all transients have died out and we  are \sout{starting from} in steady-state conditions. \sout{The solution at time $t\geq0$ is given by $x(t,X)=S^tX$ where $S^t$ is the evolution operator up to time $t$.} 
We now introduce a target function of the form $F(\tau,X)=\int_0^\tau\mathrm{d}t f(x(t,X))$ where $f$ is a smooth function of its argument. \sout{Such a functional clearly depends on the initial condition $X$ and on the final time $\tau$.} {\color{red} Such a target function can be tailored to capture the persistent extreme of interest.} We define as $\mathcal{P}(F(\tau,X)>z)$ \sout{
\begin{equation}
    P_T(z)=\mathcal{P}(F(\tau,X)>z)\label{condi}
\end{equation}}
the probability that \sout{the functional} $F(\tau,X)$ has value larger than $z$. \sout{More specifically, $\mathcal{P}(F(\tau,X)>z)=\rho(\mathbf{1}_{A_{F,\tau,z}})$, where $A_{F,\tau,z}$ is the set of points of $\Omega$ such that if $y\in A_{F,\tau,z}$, then $F(\tau,y)>z$ and $\mathbf{1}$ indicates the characteristic function. The expression $\rho(\Psi(X))=\int \rho(\mathrm{d}X)\Psi(X)$ indicates the expectation value of $\Psi$ according to the invariant measure $\rho$.} The working hypothesis is to be able to write $\mathcal{P}(F(\tau,X)>z)$ in the limit of large $z$ as a large deviation law of the form
\begin{equation}\label{LDL}
    \mathcal{P}(F(\tau,X)>z)\asymp \exp(-\min_{X\in A_{F,\tau,z}} I(X))
\end{equation}
where {\color{red} $A_{F,\tau,z}$ is the set of points of $\Omega$ such that if $y\in A_{F,\tau,z}$, then $F(\tau,y)>z$} and $\asymp$ indicates that the ratio of the logarithms of the two sides tends to 1. 
Finally, $I(X)$ is the so-called rate function. \sout{One has that  $I(X)=\max_\eta(\langle \eta,X\rangle-S(\eta))$  where $S(\eta)=\log\rho(\exp(\langle \eta,X\rangle))$ is the cumulant generating function of $X$, and $\langle,\rangle$ is the scalar product in $\mathbb{R}^N$.} As a result, one obtains that $$\mathcal{P}(F(\tau,X)>z)\asymp \exp(-I(X^*(z))),$$ where $$X^*(z)=\underset{X\in A_{F,T,z}}{\arg\min} (I(X)).$$
This implies that the most likely way the event $\mathcal{P}(F(\tau,X)>z)$ occurs is for $X=X^*(z)$. Namely, as we consider larger and larger thresholds $z$, the probability measure concentrates more and more around a specific initial condition $X^*(z)$. \sout{Note that this large deviation formulation is not the usual one because there is no large parameter controlling the size of the argument of the exponent in Eq. \ref{LDL}. Instead, the asymptotic equivalence is obtained because one considers larger and larger values of $z$.}
\begin{figure}
\includegraphics[trim=0 5cm 0 5cm, clip=true, width=1.\linewidth]{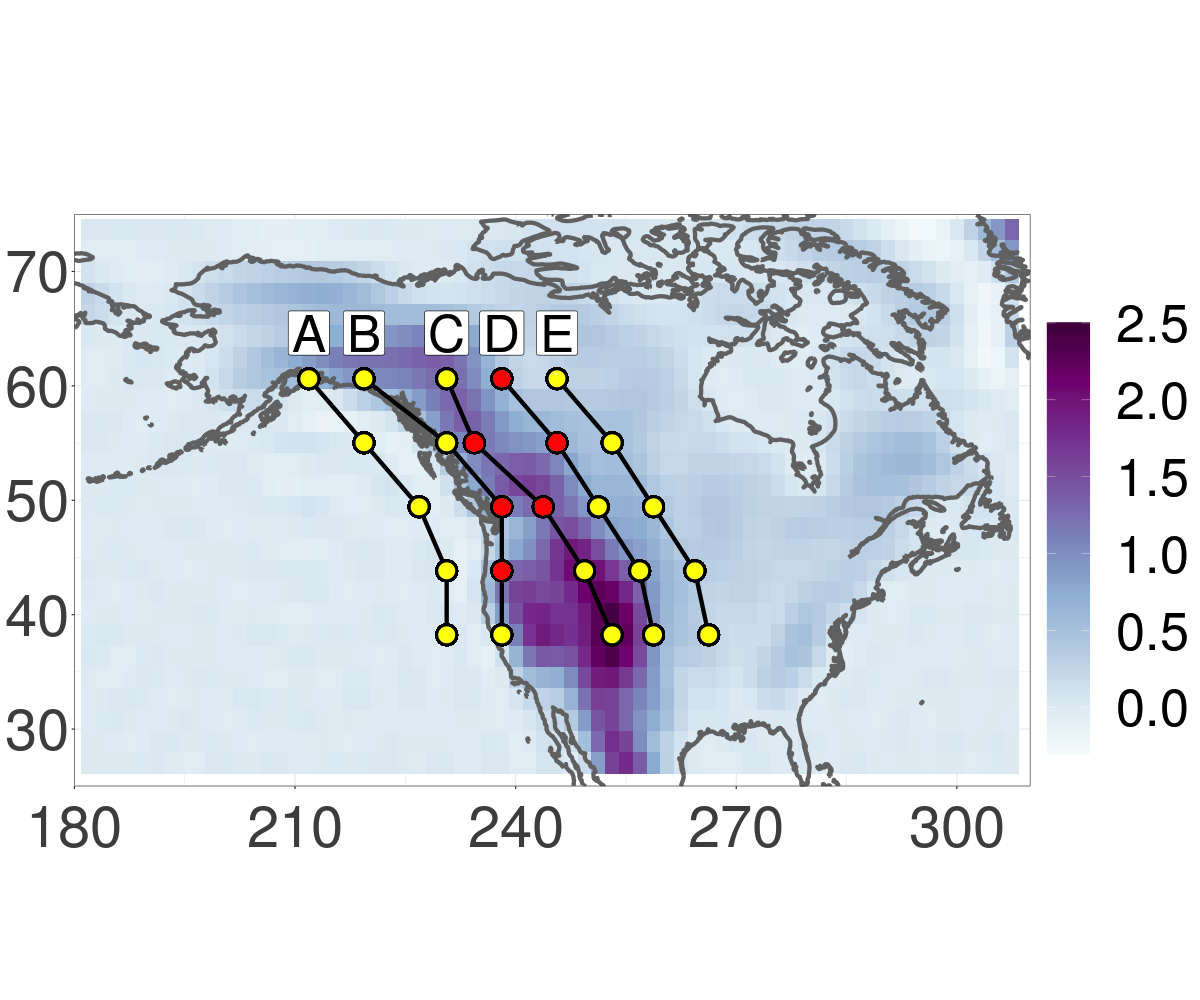}
\caption{The slanted grid shows the location of the 25 grid points considered in this study. They are arranged in an orography-following fashion and for latitudes ranging from $38^\circ N$ to $61^\circ N$. The red dots indicate the core of the 2021 WNAHW as discussed in the text. The color bar indicates the MPI-ESM-LR model orography in units of Km. 
\label{grid}}
\end{figure}
Let's provide an interpretation of this result following \cite{Touchette2015}. The initial conditions $X=X^*(z)$ around which the measure concentrates define the forward {\color{red}trajectory} $x(t,X^*(z))=S^tX^*(z)$. It is clear that such initial condition and \sout{all the ensuing orbits are} {\color{red}the ensuing forward trajectory are} extremely atypical exactly because  there is an association with the desired rare event defined in the limit of large $z$. Conversely, if one studies the statistics of the system conditional on the occurrence of the rare event {\color{red}(i.e. restricted to the set $A_{F,\tau,z}$)}, the {\color{red}trajectory} departing from $X^*(z)$ becomes typical. Indeed, {\color{red}LDT} \sout{large deviation theory} captures here the least unlikely of all the unlikely ways the rare event of interest can occur \cite{Hollander2000}. 

In {\color{red}the} previous \sout{investigations on} {\color{red}studies of} the 2010 Russian Heatwave,  2010 Mongolian Dzud (an extreme cold spell), and 2019 North American cold spell \cite{Galfi2021,Galfi2021PRL} \sout{we noted that, in each case,} the model-simulated events selected according to the procedure {\color{red} above, where $z$ was a measure of the observed extreme event at a given location in its core region}
\begin{enumerate}
    \item resembled each other: the large scale spatial patterns of {\color{red}T2M} and Z500, while being  abnormal {\color{red}in reference to} \sout{compared to} the \sout{usual} climatology, were highly correlated with the mean field computed by averaging over the various events;
    \item resembled the actual historical event: the observed {\color{red}T2M} and Z500 fields could be interpreted as being drawn from the same distribution including the various model-generated surrogate events. 
\end{enumerate} 

The properties 1. and 2. above indicate the usefulness of the mathematical framework  given in the previous section: if we consider events selected according to a sufficiently stringent constraint, they \sout{will fall in the domain of attraction of the large deviation law described above. Hence, they} can be seen as fluctuations around the optimal $x(t,X^*(z))=S^tX^*(z)$ case, which can be estimated as the average over all the selected events generated by the model \cite{Dematteis2019,DematteisSIAM2019}.



\section{Data}
We use here, on top of the T2M and Z500 ERA5 \cite{ERA5} fields for June 24th - July 8th 2021, the data for the same variables gathered from a 1000-year simulation performed by the MPI-ESM-LR {\color{red}version 1.2  Earth System Model (ESM) \cite{Mauritsen2019}} in standard pre-industrial conditions in terms of atmospheric composition and land-use dataset (\textit{piControl} run). {\color{red}This ESM is  among the best performing models participating to the  6th phase of the Climate Model Intercomparison Project (CMIP6) \cite{Eyring2016}, just  as its previous version was among the best performing CMIP5 models  \cite{Bock2020}.}

{\color{red}After removing the seasonal cycle by subtracting the climatology of each day of the year,  we select for each year an extended summer lasting $160$ days and beginning on May 5th, as in \cite{Galfi2021PRL}, in order to remove winter heatwaves from the statistics.} We consider 25  grid points located in the Western part of  North America; see Fig.~\ref{grid}. These grid points are arranged in an irregular lattice: the North/South direction is distorted in such a way that it approximately follows the orography. The grid points are defined by their latitude (38$^\circ$ N-61$^\circ$ N in steps of $\approx 6^\circ$) and by the letters A, B, C, D, and E (from west to east). A refers to points located over the ocean (except in case of latitude 61° N), not far from the coast; B to points located near the Pacific coast; C and D to points at the top and at the eastern flank of the Rocky mountains; and E to points located in the continental region {\color{red}to the East} \sout{ of  the Eastern side} of the Rocky mountains. 

\sout{The idea is to study intense and persistent heatwaves affecting such locations and to relate them to the actual 2021 WNAHW.} Following the mathematical framework described above, the first step is to use 
\sout{in Eq. \ref{condi}}
the following definition for the target function: $F_k(\tau,X)=\int_0^\tau\mathrm{d}t(T_k(t,X)-\bar{T}^\tau_k)$, where $k$ is the index referring to the  grid point of interest, $T_k(t,X)$ is the {\color{red}T2M} at time $t$ given the initial condition $X$, and $\bar{T}^\tau_k$ is the long-term climatology of $T_k$. 
During the 2021 event, for both cases of $\tau=15$ $d$ and of $\tau=7$ $d$, the target function is largest  at the following grid points (indicated in red in Fig. \ref{grid}): $44B$ (near Portland, USA), $49B$ (near Vancouver, Canada), $49C$, $55D$ (Calgary, Canada, is approximately midway between these two latter locations), $55C$, and $61D$ (near Forth Smith, Canada). See Tables \ref{delta15} and \ref{delta7}.

\section{Results}
In what follows we show that there is a clear signature of the 2021 WNAHW in the natural variability of the MPI-ESM-LR.  
\sout{In \cite{Galfi2021,Galfi2021PRL} we also noted that the large scale spatial patterns defined according to the procedure above were somewhat structurally insensitive to the selection of the reference grid point: the patterns, roughly speaking, shift following the position of the reference grid point. Since the Rocky mountains and the associated coastline are the main geographical features of the region mostly affected by the 2021 WNAHW, it is worth investigating their roles in determining its properties.}
{\color{red}We will further test whether there is evidence of a strong signature of the Rocky Mountains on the 2021 WNAHW. Indeed, orography has a key role in catalyzing and determining the location of large scale and persistent weather patterns associated with the low-frequency variability of the atmosphere  \cite{Charney1979,CharneyStraus1980,Benzi1986,Mullen1989,Malguzzi1997,Schubert2016,Ruti2006,Narinesingh2020} and in determining the properties of heatwaves \cite{Jimenez2022,Jimenez2022b}.}
\sout{In particular, one should take into account the great importance of}

The 1000-year MPI-ESM-LR model run features by and large events comparable to the 2021 WNAHW {\color{red} in terms of observed anomalies.} The return times for values of the target function at the locations indicated above range from multidecadal to multicentennial. In the case $\tau=15$ $d$, they reach a value of  200 $y$ at grid point $49C$ and at 91 $y$ at grid point $55D$. In the case $\tau=7$ $d$, one gets multicentennial values at grid points $49B$ and $49C$. Instead, the model cannot find an event of intensity equal or larger than the 2021 WNAHW for grid point $55D$. Comparing Tables \ref{delta15} and \ref{delta7} one {\color{red}finds confirmation of the presence of a very intense phase within the heatwave}, as \sout{. appreciates the relatively short duration of the event,} already mentioned in the Introduction.
For the other grid points shown in Fig. \ref{grid} and not included in Tables \ref{delta15} and \ref{delta7}, instead, the return times of the 2021 value of $F_k(\tau,X)$ estimated through the MPI-ESM-LR model dataset are at most of the order of few years. Hence, they are relatively (or very) common within, e.g., a 30-year time span commonly used to benchmark climatology. We conclude that  the red grid points in Fig. \ref{grid} define the (vast) core region of the 2021 event, which results to be larger than the area investigated in \cite{Philip2021,Thompson2022}. 

\sout{Note  that, when using the model data, we consider only data relative to the extended summer season which, as in \cite{Galfi2021PRL}, ranges from May to mid October, in order to remove winter heatwaves from the statistics.} 


\paragraph{Robustness of the Methodology}
\sout{It is important to first verify} {\color{red}We first test} whether the mathematical framework presented above provides us with a robust and practically usable methodology. We proceed as follows. As shown in Tables \ref{delta15} and \ref{delta7}, the 50-year return level estimated by the model is  by and large a good \sout{ballpark} {\color{red}reference} value for the observed anomalies for all of the grid points of the core region of the 2021 WNAHW. {\color{red}The results are weakly dependent on the specific choice of the return time, as long as it is sufficiently long; see discussion below.} We then select, for each of the 25 grid points shown in Fig. \ref{grid}, as threshold $z_k$ for $F_k(\tau,X)$ the 50-year return level. In such a way, we have a homogeneous statistics throughout the domain of interest for extreme heatwaves, because we identify $J=20$ events for each of the 25 grid points of interest. Each event corresponds to one initial condition $X^j$ compatible with $F_k(\tau,X)>z_k$. {\color{red}In other terms, our procedure aims at defining candidates for weather analogues \cite{Lorenz1969,Zorita1999,Vautard2009} for the  extreme event  of interest. } \sout{In what follows w}{\color{red}We present below} results for $\tau=15$ $d$. All the main conclusions we draw are valid also for $\tau=7$ $d$, whose corresponding figures are shown in the SM; see further discussion therein.

\begin{table}
\begin{tabular}{ |c|c|c|c|c|c| } 
\hline
Grid point & $RL_{50}$ & $\Delta {\color{red}T2M}$ & $RT(\Delta {\color{red}T2M})$ & $RT(\Delta {\color{red}T2M}-\delta )$  \\
\hline
44B & 6.3 & 5.6 & 19 & 6  \\ 
49B & 6.0 & 5.4 & 22 & 5  \\ 
49C & 5.6 & 6.2 & 200 & 22  \\
55C & 7.3 & 6.7 & 29 & 8  \\
55D & 5.2 & 5.5 & 91 & 17  \\
61D & 5.7 & 5.3 & 30 & 6  \\
\hline
\end{tabular}
\caption{Surface temperature anomalies ($\Delta {\color{red}T2M}$, in $^\circ C$) averaged over $\tau=15$ $d$ for the 2021 WNAHW at selected grid points (see Fig. \ref{grid}) in its core region (third column). For each grid point, we have $\Delta {\color{red}T2M}=F_k(\tau,X)/\tau$. The corresponding return times (RTs, in $y$) estimated with the MPI-ESM model (fourth column) as well as the the RTs for the  climatology shifted by $\delta=1.2^\circ$ $C$ (last column) are reported. The model return levels for 50-year return times ($RL_{50}$, in $^\circ C$) are reported in the second column.\label{delta15}}
\end{table}

\begin{table}
\begin{tabular}{ |c|c|c|c|c| } 
\hline
grid point & $RL_{50}$ & $\Delta {\color{red}T2M}$ & $RT(\Delta {\color{red}T2M})$ & $RT(\Delta {\color{red}T2M}-\delta )$  \\
\hline
44B & 8.8 & 8.2 & 29 & 8   \\ 
49B & 8.4 & 9.7 & 333 & 53  \\ 
49C & 7.6 & 8.9 & 500 & 62  \\
55C & 10 & 9 & 21 & 9  \\
55D & 7.2 & 9.1 & $>1000$ & 1000  \\
61D & 7.9 & 7.5 & 36 & 9\\
\hline
\end{tabular}
\caption{Same as Table \ref{delta15}, but for $\tau=7$ $d$.\label{delta7}}
\end{table}


We first test whether by considering for a given $k$ such a high threshold $z_k$ for the function $F_k(\tau,X)$ we end up identifying sufficiently similar weather patterns, which can be seen as fluctuations around $X^*(z)$. \sout{, as defined in Eq. \ref{LDL}.} This amounts to being able to introduce a notion of typicality for such extreme events {\color{red} that leads to identifying good weather analogues.}

\begin{figure*}\includegraphics[width=0.9\linewidth]{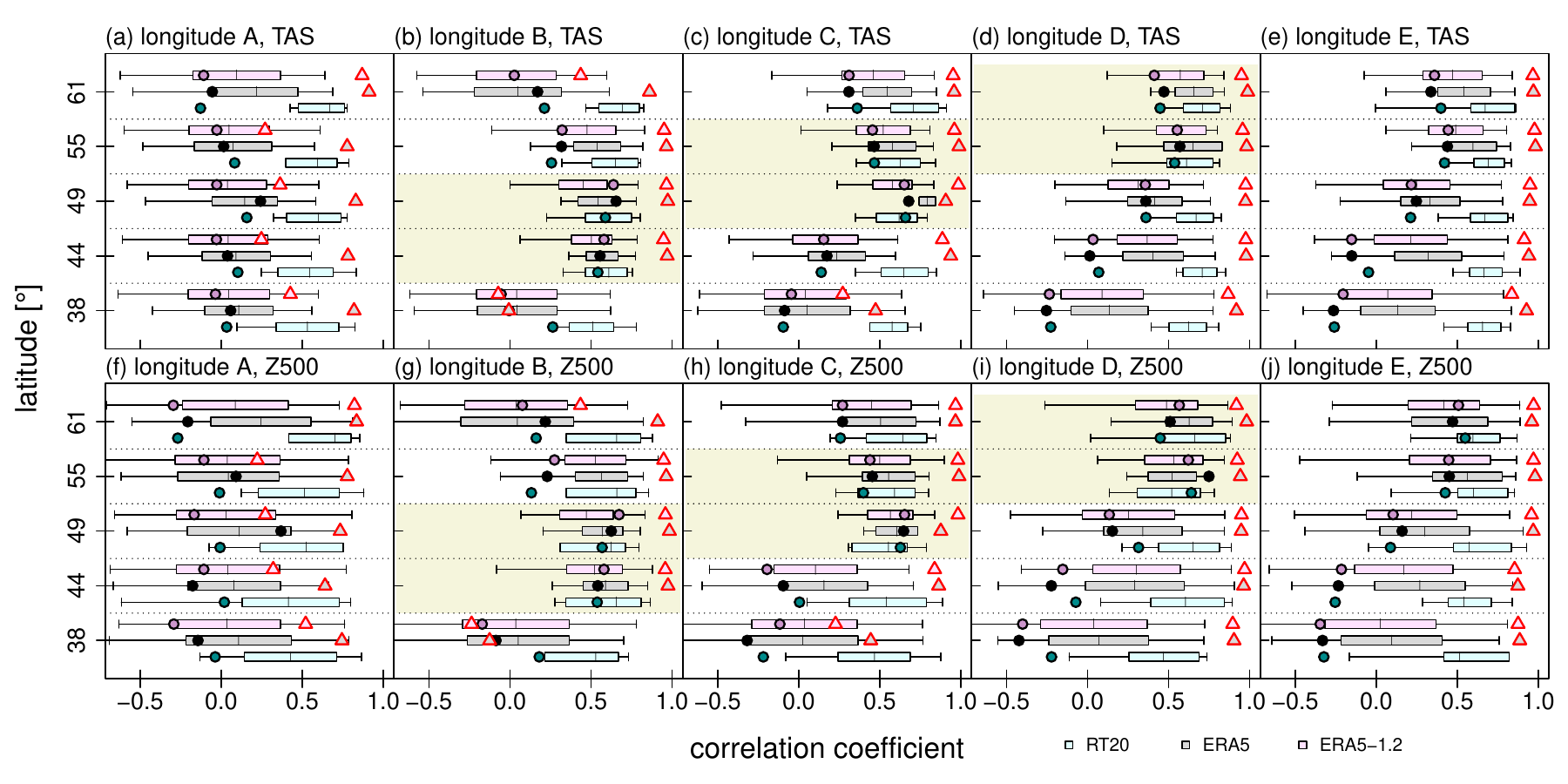}\caption{Spatial correlation coefficients between the model-simulated events and their mean for the {\color{red}T2M} (a-e) and Z500 (f-j) fields. The coefficients are obtained for events selected by defining a high threshold in the cumulated temperature anomaly at the corresponding grid point over $\tau=15$ $d$; see indication of the latitude on the side and of the longitude on top. The shading indicates the core region of the 2021 WNAHW, see Tables \ref{delta15}-\ref{delta7}. In each subpanel, the boxes refer to the $\pm 1\sigma$ interval, and the whiskers indicate the full range of results. The dots indicate the spatial correlation coefficient between the mean of the model runs and the corresponding field from ERA5. Additionally, i) Cyan refers to 50-y return time events, ii) Gray refers to events with return level corresponding to the one of the 2021 WNAHW,  and iii) Magenta refers  to events where the return level of 2021 WNAHW event is reduced by $\delta=1.2^\circ C$. 
The triangles indicate the spatial coefficient between the mean of the events i) and the mean of the events ii) and iii), respectively. {\color{red}See the Taylor diagrams in Figs. S3-S14 in the SM for more complete information on the agreement between the considered patterns.} \label{coherencerealism}}
\end{figure*}

The cyan boxplots in Fig. \ref{coherencerealism} show the properties of the spatial correlation between the $\tau$ averages of {\color{red}T2M} and Z500 fields of the $J=20$ events selected for each grid point $k$ and their respective mean. Since we are now testing the robustness of the methodology, we perform this analysis for all the grid points shown in Fig. \ref{grid}.  

The spatial correlation is computed over the region [20$^\circ$ N, 75$^\circ$ N]$\times$[150$^\circ$ E, 300$^\circ $E], which includes the whole North America, approximately half of the North Pacific ocean, and a quarter of the North Atlantic ocean. The results shown below are very weakly sensitive to adjustments in the boundaries of such a (vast) region. We clearly see that in most cases the mean of spatial correlation is high, with a moderate standard deviation, which indicates high coherence between the events. In general, the coherence is higher as we focus on higher latitudes, thus reducing the impact of tropical dynamics, and as we move away from the ocean towards the Rockies, in agreement with what is mentioned above regarding the interaction between orography and low-frequency (and large scale) mid-latitude atmospheric variability. 


{\color{red}Stronger support to these statements can be found by looking at the Taylor \cite{Taylor2001} diagrams shown in the SM as Figs. S3-S4 (S5-S6) for the T2M (Z500) field, 15-day and 7-day averages, respectively. Taylor diagrams allow for a more complete comparison between fields by including, on top of their spatial correlations, also the centered root-mean-square (RMS) difference and the amplitude of their variations (represented by their standard deviations). For most of the 25 grid points of interest, the 20 50-y return level events define a cluster - see below - characterised by high correlation and low centered RMS difference with respect to the mean of such events. 

We remark that for the grid points over the ocean (where the correlation of T2M has a slower decay) and in case of the southernmost latitude (where there is strong interaction between tropical and extratropical dynamics), the spread is substantially larger because it is more difficult to obtain a well-defined typicality there \cite{Galfi2021PRL}.

Clearly, because of averaging, the mean tends to have lower standard deviation than the individual events. As we consider longer and longer return times - see Figs. S7-S10 in the SM - the clusters become better defined as the individual events become more and more similar to their averages.  Indeed, in each Taylor diagram the center of the cluster moves in the direction of the reference point in the x-axis. The presence of a single well-defined statistical cluster for the T2M and Z500 fields (separately and taken together) is confirmed in all cases for fields constructed for grid points over land and for return times longer than 20 years through the gap statistics \cite{Tibshirani2001} for Gaussian mixture models \cite{McLachlan2000}, which has been evaluated using the MATLAB  function \texttt{evalclusters} \cite{MATLAB:2020}.}


We conclude that the methodology followed here allows us to distill \sout{reasonably well} the fundamental properties of heatwaves throughout the domain of interest. \sout{We remark that t}{\color{red}T}he test above is not complete because, for each $k$, one should look at the agreement between the evolution up to time $\tau$ of the various $j=1,\ldots J$ initial conditions $X^j$. Instead, here we are looking at the agreement between the fields $\bar{X}^j_\tau=1/\tau\int_0^\tau\mathrm{d}tS^tX^j$, and we estimate $\bar{X}^*(z)_\tau=1/J\sum_{l=1}^J \bar{X^j}_\tau$. Nonetheless, the results we obtain are extremely encouraging.

\paragraph{Fingerprinting the 2021 WNAHW}

The next step is to discover whether there is an agreement between the heatwaves constructed as above and the 2021 WNAHW. Indeed, if we repeat the analysis  by using the actual return levels of the 2021 WNAHW - see gray boxplots in Fig. \ref{coherencerealism} - we obtain for the grid points in the core region results that are in  \sout{excellent}{\color{red}very good} agreement with what found using the 50-y return levels. {\color{red}This confirms that the} estimate of $X^*(z)$ depends weakly on $z_k$ as long as we are choosing a sufficiently large value for $z_k$. 

\begin{figure*}
(a)\includegraphics[trim=42cm 25cm 127cm 77cm, clip=true, width=0.42\linewidth]{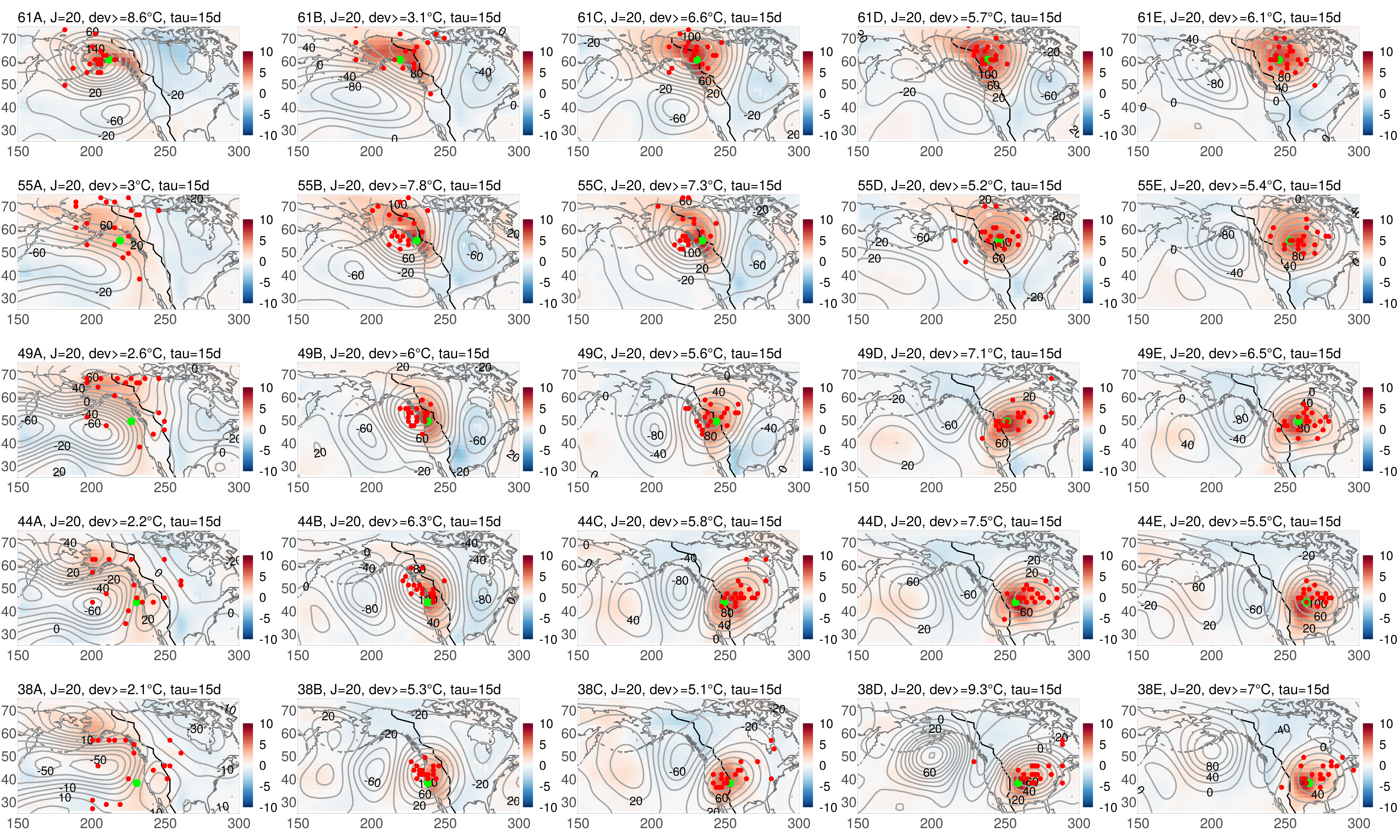}
(b)\includegraphics[trim=42cm 50.3cm 127cm 51.7cm, clip=true, width=0.42\linewidth]{./tas_zg_gp_n15_nevents_20_model_des_nh_maxpoints_maxpointsocean_smallbox_smallermap_allpoints_row.png}\\
(c)\includegraphics[trim=84.3cm 50.3cm 84.7cm 51.7cm, clip=true, width=0.42\linewidth]{./tas_zg_gp_n15_nevents_20_model_des_nh_maxpoints_maxpointsocean_smallbox_smallermap_allpoints_row.png}
(d)\includegraphics[trim=84.3cm 75.6cm 84.7cm 26.4cm, clip=true, width=0.42\linewidth]{./tas_zg_gp_n15_nevents_20_model_des_nh_maxpoints_maxpointsocean_smallbox_smallermap_allpoints_row.png}\\
(e)\includegraphics[trim=126.6cm 75.6cm 42.4cm 26.4cm, clip=true, width=0.42\linewidth]{./tas_zg_gp_n15_nevents_20_model_des_nh_maxpoints_maxpointsocean_smallbox_smallermap_allpoints_row.png}
(f)\includegraphics[trim=126.6cm 100.9cm 42.4cm 1.1cm, clip=true, width=0.42\linewidth]{./tas_zg_gp_n15_nevents_20_model_des_nh_maxpoints_maxpointsocean_smallbox_smallermap_allpoints_row.png}
\caption{{\color{red}T2M} and Z500 anomaly fields (averaged over $\tau=15$ d) computed as means over the $J=20$ most extreme model-simulated events chosen according to a high {\color{red}T2M} anomaly thresholds at grid points $44B$ (a), $49B$ (b), $49C$ (c), $55C$ (d), $55D$ (e) and $61D$ (f). The red dots indicate the center of the high pressure for each of the $J$ events. All the patterns look remarkably similar, and similar to the 2021 WNAHW, see Fig. \ref{anomaly7}.}\label{patternsLDL}
\end{figure*}

The full dots in the gray boxplots in Fig. \ref{coherencerealism} show, for each $k$, the value of the spatial correlation coefficient between the estimate of  $\bar{X}^*(z)_\tau$ and the actual $\tau$-averaged ERA5 fields depicted in Fig. \ref{anomaly7}. The spatial correlation coefficient for both the {\color{red}T2M} and Z500 field is highest when considering the events constructed for the grid points defined in Tables \ref{delta15} and \ref{delta7}.   
For five out of these six grid points, additionally, we have that the value of the spatial correlation coefficient is within the range of values obtained between $\bar{X}^j_\tau$ and $\bar{X}^*(z)_\tau$, and in most cases within the one standard deviation interval. Hence, the ERA5 anomaly field can be seen as belonging to the same statistical ensemble as the anomalies produced by the model, given the stringent criteria set for suitably choosing the corresponding events. The only exception comes from the grid point $49C$, basically because we have only few model occurrences of heatwaves of larger intensity than the 2021 WNAHW. Such events closely resemble each other, so that the range of values of correlations with their average is very small. Nonetheless, the  spatial correlation between the estimated $\bar{X}^*(z)_\tau$ and the actual $\tau$-averaged ERA5 is \sout{extremely} high. {\color{red}The Taylor diagrams shown in Fig. S11 (T2M) and Fig. S13 (Z500) of the SM further support these conclusions. Similar results are obtained when considering $\tau=7$ $d$ (see Figs. S12 and S14 of the SM), except for the fact that the ensemble is too limited to draw conclusions for the grid points $49B$ and $49C$, and no ensemble members are available for the grid point $55D$, possibly as a result of the model's inability to reproduce specific local processes.}  



The similarity of the six panels of Fig. \ref{patternsLDL} among themselves, with Fig. \ref{anomaly7}b, {\color{red}and with the corresponding panels of Fig. S16 of the SM (which are constructed using the actual return levels of the 2021 WNAHW)} provides a visual confirmation of what mentioned above. In meteorological terms, the possibly most striking common features between the model data and the actual 2021 WNAHW are a) the presence of a wavetrain across the Pacific ocean \cite{Lin2022}, which closely resembles the anomaly pattern deemed responsible for drought over California \cite{Teng2017}; and b) the powerful ridge in the mid-latitudes extending from the Pacific over the Rockies, which reminds of the  pattern that which is  associated with the occurrence of intense drought in the Western US \cite{Seager2014,Swain2015,Teng2017,Zhang2022}.

We are now able to strengthen the statement made before regarding the fact that the grid points included in Tables \ref{delta15} and \ref{delta7} correspond to the core of  the 2021 WNAHW, which was based purely on the intensity of the local T2M anomalies. \sout{Arguably, f}{\color{red}F}or these six grid points we have both high internal coherence among the model-generated events and high statistical compatibility with the large-scale fields of the 2021 WNAHW. Geographically, these grid points correspond to  a region stretching approximately in the SSW-NNE direction and, hence, shifting from the coastal area to the eastern side of the Rockies as one progresses northwards. \sout{We conclude that t}{\color{red}T}hese  grid points define a coherent region associated with a very special yet very robust feature of the natural variability of the climate, which most recently manifested itself as the 2021 WNAHW. As can be seen by comparing the corresponding panels of Figs. S2 and S16, for each of these grid points the average pattern of the model-generated events is virtually unchanged if we consider either a) the 20  events leading to the largest local persistent temperature anomaly (also portrayed in Fig. \ref{patternsLDL}) or b) events leading to local persistent temperature anomalies equal or larger than what observed during the 2021 WNAHW. Such an agreement can also be checked by noting that for such grid points the triangles next to the grey boxes in Fig. \ref{coherencerealism} are located very close to one. 

Let's  look at the grid points outside the 2021 WNAHW core region. The gray boxplots, constructed according to the 2021 WNAHW return levels, cover a larger range of values and are substantially shifted towards lower values compared to the corresponding cyan ones. This means that if we select soft extremes (or events that are not extreme at all), the notion of typicality discussed above is lost, because we are outside the regime where LDT applies. \sout{The spatial correlation coefficient} The Taylor diagrams indicate a much reduced agreement between the average of the  model generated {\color{red}T2M} and Z500 fields and the corresponding ERA5 fields \sout{is in general much lower than for}{\color{red}as compared to the case of} the core region.

Specifically, from Fig. \ref{coherencerealism} {\color{red}and Figs. S3-S6 of the SM, one can see that a good agreement still exists while considering averages constructed for grid points to the NNE of the core region, such as $55E$ and $61E$. Instead, the correlation dramatically decreases when considering grid points geographical close to the core region, such as $44C$ and $55B$, but  off-axis with respect to the SSW-NNE direction. Hence, at least in the MPI-ESM-LR model world, heatwaves centred in, e.g., grid points $44C$ and $55B$ are due to dynamical processes that are fundamentally different \sout{to} {\color{red}from} those responsible for the 2021 WNAHW. 

\section{Discussion and Conclusions}
The 2021 WNAHW  shocked experts and general audiences around the world and had devastating impacts on a large swath of land in North America. We have investigated this event  following the strategy delineated in \cite{Galfi2021PRL,Galfi2021} and based upon the theoretical framework defined in \cite{Touchette2015,DematteisSIAM2019}. While the 2021 WNAHW is very rare and unusual according to the overall statistics of weather variability, it becomes \textit{typical} when considering the statistics conditional on the validity of a  specific constraint, namely the occurrence of a large and persistent summer temperature anomaly at one specific location, defined by a grid point of our model. Typicality results, against intuition, exactly from \textit{the extreme intensity} of the event. Indeed, we find a good agreement, {\color{red}as shown by Fig. \ref{coherencerealism} and by the Taylor diagrams included in the SM}, between the large scale meteorological anomaly patterns of the 2021 WNAHW and those of the events selected from the 1000-year-long MPI-ESM-LR preindustrial run according to the procedure above when \sout{selecting}{\color{red}considering} any grid point located in a specific region of Western North America. This region has an extension of 2000 Km in the SSW-NNE direction and defines the  core  of the 2021 event. 

This indicates that the circulation pattern causing heatwaves over this region in the model resembles the one that caused the 2021 WNAHW, at least in the spatial distribution of mid-tropospheric and surface anomalies. This also indicates that, at least in the model climate, there is, roughly speaking, mainly one way to get heatwaves of such large magnitude in that region. 
\sout{We remark that this angle allows us to better frame a holistic view of the heatwave as a large-scale event constraining the events on extreme anomalies occurring at specific locations it may be possible to find in the observational data temperature deviations that are not reproduced by the model, as in the case of $\tau=7$ $d$ for the grid point $55D$, possibly because of very local features. We envision the possibility of addressing this issue by performing  spatial coarse graining of the temperature fields.} 

The obtained large scale patterns are in good correspondence with those deemed responsible for the \sout{dry anomalies leading to} persistent drought in the western US. {\color{red}This supports the existence of a cascading process} {\sout{strengthening  feedback} acting across different {\color{red}spatial and} temporal scales that is increasing dramatically climate risk 
in the western sector} of the North American continent, with conditions being and becoming particularly exacerbated in California \cite{AghaKouchak2014,Diffenbaugh2015,Mann2015}.

 Modulating the imposed threshold on the local {\color{red}T2M} anomaly amounts to taking into account, in first approximation, changing background climatic conditions. Shifting the mean has been recently shown to represent a good first approximation for evaluating the statistics of hot temperature extremes in a changing climate \cite{Gessner2021,Thompson2022}. This can be explained - in the case of persistent extremes - by the fact that the rate function for summer (as opposed to winter) temperature near the region of interest seems to be approximately insensitive to the levels of $CO_2$ \cite{Galfi2021PRL}. We find - see Tables \ref{delta15}-\ref{delta7} - that in the core of the 2021 WNAHW the return times of events comparable to the 2021 event decrease substantially (by a factor of about 3 to 10) if one budgets in an offset ($\delta=1.2^\circ$ $C$) that approximately represents the historical shift of temperatures due to the realized climate change between pre-industrial conditions and today. 
 This \sout{clearly} indicates that the likelihood of {\color{red}events comparable to the 2021 WNAHW} is greatly enhanced by climate change: {\color{red}a} \sout{the} very or extremely unusual {\color{red}event} becomes comparably likelier. Interestingly, in the core region the typicality of the extreme events is maintained even if we relax a bit our procedure: the large scale meteorological patterns associated with events selected according to the less stringent (because of the offset) constraint imposed in the core region are in good agreement with those obtained using the more stringent criterion; see the magenta and gray boxplots in Fig. \ref{coherencerealism}. 
 
\sout{We argue that the 2021 heatwave was \sout{not a \textit{Dragon King} but rather} an unlikely manifestation of climate variability. }The mathematical setting behind  this paper indicates why dynamical similarity exists between the various extreme events, despite their different intensity, as long as they are all chosen according to the same sufficiently stringent constraint. Our analysis is limited in scope by the fact that we look at the data produced by only one state-of-the-art ESM {\color{red} - yet a model with rather good overall skill as compared to the others included in the CMIP6 exercise \cite{Eyring2016}}, and at least some of our conclusions could be affected by its biases and deficiencies. It would be important to test whether similar results can be found in other CMIP6 models, also taking into account that ESMs tend to globally underestimate duration, intensity, and frequency of heatwaves, yet having a comparatively better performance in the West North America region \cite{Hirsch2021}. Interestingly, the CMIP6 bias is instead on the warm side for daily temperature maxima \cite{DiLuca2020}, which further reinforces the need to deal carefully with the property of persistence when looking at hot events. The methodology  adopted here could be replicated for studying the statistics and dynamics of several historical extreme heatwaves, taking advantage of the dataset compiled by \cite{Thompson2022}. 

In order to better characterize the  features of the event studied here - with the goal, e.g., of better understanding the separate role of dynamical and thermodynamical contributions,  preconditioning factors, and the role of teleconnections - one would need larger statistical sets than those considered here, because we construct our ensembles by imposing rather stringent conditions on the fields. {\color{red}A possible strategy is to take advantage of the  CMIP6 dataset  \cite{Eyring2016}, and use all available ensemble members and various climate scenarios in order to address the problem of the impact of climate change on the statistics and dynamics of heatwaves. A theoretically more challenging yet more promising strategy is, instead, to exploit rare events algorithms to disproportionally populate the exotic configurations of interest
 and learn a great deal more about them \cite{Ragone2017,RagoneBouchet2021}. Nonetheless, both of these directions are beyond the scopes of this investigation as they require extensive data analysis and modelling work.}

{\color{red}Our approach can also be seen as leading to the definition of weather analogues for extreme events.} In \cite{Faranda2021} it was shown, by investigating the recurrence of weather analogues, that in the last decades there has been a positive trend in the frequency of occurrence of atmospheric circulation patterns over the North Atlantic  driving summertime dryness and heatwaves and a negative trend in those leading to  wet, cooler summer conditions across Europe. We envision applying the methodology proposed in \cite{Faranda2021} to the West North America region \sout{in order} to complement the analysis presented here in the direction of understanding the dynamical properties responsible for the exacerbating dry conditions in the region.

\paragraph{Acknowledgments}
The authors thank D. Faranda, T. Grafke, F. Ragone, J. Wouters for many exchanges on extreme events, L. Recchia for her comments on a preliminary version of the manuscript, {\color{red} and two anonymous reviewers for constructive criticism}. VL acknowledges the support by the Horizon 2020 project TiPES (grant no. 820970), by the Marie Curie ITN CriticalEarth (grant no. 956170) and by the EPSRC project EP/T018178/1. GM and JR acknowledge the European Research Council (ERC) under the European Union's Horizon 2020 research and innovation programme (project CEN\AE: compound Climate Extremes in North America and Europe: from dynamics to predictability, grant no. 948309). VMG acknowledges the support of the {\color{red}Air, Water and Landscape Science} Programme at the Department of Earth Sciences of Uppsala University.




\bibliographystyle{abbrv}
\bibliography{apssamp}


\end{document}